\documentstyle[epsfig]{mn}

\newif\ifAMStwofonts
\AMStwofontstrue

\ifoldfss
  \newcommand{\rmn}[1] {{\rm #1}}

  \ifCUPmtlplainloaded \else
    \NewTextAlphabet{textbfit} {cmbxti10} {}
    \NewTextAlphabet{textbfss} {cmssbx10} {}
    \NewMathAlphabet{mathbfit} {cmbxti10} {} 
    \NewMathAlphabet{mathbfss} {cmssbx10} {} 
  \fi
  \ifAMStwofonts
    \ifCUPmtlplainloaded \else
      \NewSymbolFont{upmath} {eurm10}
      \NewSymbolFont{AMSa} {msam10}
      \NewMathSymbol{\upi}     {0}{upmath}{19}
      \NewMathSymbol{\umu}     {0}{upmath}{16}
      \NewMathSymbol{\upartial}{0}{upmath}{40}
      \NewMathSymbol{\leqslant}{3}{AMSa}{36}
      \NewMathSymbol{\geqslant}{3}{AMSa}{3E}

      \let\geq=\geqslant 
    \fi
  \fi
\fi 

\ifnfssone
  \newmathalphabet{\mathit}
  \addtoversion{normal}{\mathit}{cmr}{m}{it}
  \addtoversion{bold}{\mathit}{cmr}{bx}{it}
  \newcommand{\rmn}[1] {\mathrm{#1}}

  \newmathalphabet{\mathbfit} 
  \addtoversion{normal}{\mathbfit}{cmr}{bx}{it}
  \addtoversion{bold}{\mathbfit}{cmr}{bx}{it}
  \newmathalphabet{\mathbfss} 
  \addtoversion{normal}{\mathbfss}{cmss}{bx}{n}
  \addtoversion{bold}{\mathbfss}{cmss}{bx}{n}
  \ifAMStwofonts
    \ifCUPmtlplainloaded \else
      \UseAMStwoboldmath
      \makeatletter
      \new@mathgroup\upmath@group
      \define@mathgroup\mv@normal\upmath@group{eur}{m}{n}
      \define@mathgroup\mv@bold\upmath@group{eur}{b}{n}
      \edef\UPM{\hexnumber\upmath@group}
      \new@mathgroup\amsa@group
      \define@mathgroup\mv@normal\amsa@group{msa}{m}{n}
      \define@mathgroup\mv@bold\amsa@group{msa}{m}{n}
      \edef\AMSa{\hexnumber\amsa@group}
      \makeatother
      \mathchardef\upi="0\UPM19
      \mathchardef\umu="0\UPM16
      \mathchardef\upartial="0\UPM40
      \mathchardef\leqslant="3\AMSa36
      \mathchardef\geqslant="3\AMSa3E

      \let\geq=\geqslant 
    \fi
  \fi
\fi 

\ifnfsstwo
  \newcommand{\rmn}[1] {\mathrm{#1}}

  \DeclareMathAlphabet{\mathbfit}{OT1}{cmr}{bx}{it}
  \SetMathAlphabet\mathbfit{bold}{OT1}{cmr}{bx}{it}
  \DeclareMathAlphabet{\mathbfss}{OT1}{cmss}{bx}{n}
  \SetMathAlphabet\mathbfss{bold}{OT1}{cmss}{bx}{n}
  \ifAMStwofonts
    \ifCUPmtlplainloaded \else
      \DeclareSymbolFont{UPM}{U}{eur}{m}{n}
      \SetSymbolFont{UPM}{bold}{U}{eur}{b}{n}
      \DeclareSymbolFont{AMSa}{U}{msa}{m}{n}
      \DeclareMathSymbol{\upi}{0}{UPM}{"19}
      \DeclareMathSymbol{\umu}{0}{UPM}{"16}
      \DeclareMathSymbol{\upartial}{0}{UPM}{"40}
      \DeclareMathSymbol{\leqslant}{3}{AMSa}{"36}
      \DeclareMathSymbol{\geqslant}{3}{AMSa}{"3E}

      \let\geq=\geqslant 
    \fi
  \fi
\fi 

\ifCUPmtlplainloaded \else
  \ifAMStwofonts \else 
    \def\upi{\pi}
    \def\umu{\mu}
    \def\upartial{\partial}
  \fi
\fi

\newcommand{\te}{T_{\rmn{e}}}
\newcommand{\tr}{T_{\rmn{r}}}
\newcommand{\me}{m_{\rmn{e}}}
\newcommand{\thte}{\Theta_{\rmn{e}}}
\newcommand{\kb}{k_{\rmn{B}}}
\newcommand{\ud}{{\rmn{d}}}

\newcommand{\sigt}{\sigma_{\rmn{T}}}
\newcommand{\clp}{{\mathcal{P}}}
\newcommand{\clh}{{\mathcal{H}}}
\newcommand{\cll}{{\mathcal{L}}}
\newcommand{\nelect}{\tilde{n}_{\rmn{e}}}
\newcommand{\nelec}{n_{\rmn{e}}}
\newcommand{\cle}{{\mathcal{E}}}
\newcommand{\vt}{V_{\rmn{t}}}

\title[CMB polarization from galaxy clusters]{Thermal and kinematic
corrections to the microwave background polarization induced by galaxy
clusters\\%
along the line of sight}
\author[A. D.~Challinor, M. T.~Ford and A. N.~Lasenby]
{A. D.~Challinor\thanks{Email: A.D.Challinor@mrao.cam.ac.uk},
M. T.~Ford and A. N.~Lasenby\\
Astrophysics Group, Cavendish Laboratory, Madingley Road, Cambridge,
CB3 0HE, UK}
\date{\today}

\begin{document}

\maketitle

\begin{abstract}
We derive analytic expressions for the leading-order corrections to the
polarization induced in the cosmic microwave background (CMB) due to
scattering off hot electrons in galaxy clusters along the line of sight.
For a thermal distribution of electrons with kinetic temperature
$\kb\te\sim 10\,\rmn{keV}$ and bulk peculiar velocity
$V \sim 1000\,\rmn{kms^{-1}}$, the dominant corrections to the
polarization induced by the primordial CMB quadrupole and the cluster
peculiar velocity arise from electron thermal motion and are at the level
of $\sim 10$ per cent in each case, near the peak of the
polarization signal. When more sensitive measurements become feasible,
these effects will be significant for the determination
of transverse peculiar velocities, and the value of the CMB quadrupole at
the cluster redshift, via the cluster polarization route.
\end{abstract}


\section{Introduction}
\label{intro}

Compton scattering of cosmic microwave background photons off electrons
in ionised intra-cluster gas leads to intensity distortions along the line of
sight (Sunyaev \& Zel'dovich 1972, 1980).
The same scattering mechanism will also change the
polarization state of the CMB if the radiation has a quadrupolar anisotropy
in the average rest frame of the electrons. Such a quadrupole can arise
from several physical effects, including: (i) transformation of a primordial
quadrupole moment in the frame in which the CMB dipole vanishes
(hereafter referred to as the CMB frame); (ii) aberration and Doppler
effects in the transformation of the non-quadrupole moments of the
primordial anisotropy due to electron bulk
velocities; (iii) anisotropic scattering in the cluster; (iv) gravitational
lensing from dynamic structures. The first three of these mechanisms were
originally studied by Sunyaev \&\ Zel'dovich~\cite{sunyaev80,zeldovich80},
while the fourth was considered by Gibilisco~\shortcite{gibilisco97}.

Although the small levels of polarization expected from cluster effects
fall below current experimental sensitivities, the prospect of detection
with future CMB experiments, such as the
\emph{Planck Surveyor} satellite, has fuelled a recent rise in interest
in the topic. Audit \&\ Simmons~\shortcite{audit99} have (numerically)
computed the frequency dependence of the polarization arising from the
transformation of the CMB monopole for a cold cluster, and Itoh, Nozawa
\&\ Kohyama~\shortcite{itoh99} derived an analytic expression for the frequency
dependence
of this polarization mechanism, including the first two corrections arising
from thermal effects. The claims from these two groups that the original
results of Sunyaev \&\ Zel'dovich were in error were met with a paper from
Sazonov \&\ Sunyaev~\shortcite{sazonov99}, who confirmed the earlier
Sunyaev--Zel'dovich
analysis, and derived analytic results for the frequency dependence of the
polarization arising from the CMB monopole and primordial quadrupole
in the limit of a cold electron gas. Sazonov \&\ Sunyaev~\shortcite{sazonov99}
also discussed the polarization effect due to prior anisotropic scattering
(which is second-order in the optical depth, $\tau$).

In this paper we reconsider the question of polarization generation from
clusters, in the limit where second-order optical depth and
gravitational lensing effects can be ignored. We use a convenient, covariant
representation of the linear polarization, which simplifies the
geometric interpretation of our results, to derive
simple analytic expressions for the frequency dependence of the polarization
signal. Our expressions include leading-order thermal and kinematic
corrections to the results derived by Sazonov \&\
Sunyaev~\shortcite{sazonov99}, and confirm their results in the limit of low
electron temperatures and
bulk velocities. Our results for the thermal corrections to the polarization
arising from the electron bulk velocity differ significantly from those
given by Itoh et al.~\shortcite{itoh99}. We show that for typical cluster
parameters, the leading-order corrections
to the polarization from the primordial quadrupole and the electron
bulk velocity amount to reductions in these effects of $\sim 10$ per cent
at the peak of their signals.
These corrections should thus be considered in any future
analysis of forthcoming CMB polarization data in the directions of clusters,
when such detections are permitted by improvements in receiver sensitivity.

The paper is arranged as follows. In Section~\ref{scatter} we outline our
covariant procedure for calculating the scattering term, which appears in the
the Boltzmann equation, for an unpolarized, but otherwise
arbitrary, radiation field in the CMB frame. In Section~\ref{results}, we
present our results for the polarization signal arising from the individual
multipoles of the primordial CMB intensity, and discuss their relative
importance for a typical cluster. We end in Section~\ref{conc} with our
conclusions. We employ units with $c=h=1$ throughout, and a $+---$ signature
for the spacetime metric.

\section{Polarization due to electron scattering of an
anisotropic radiation field}
\label{scatter}

We consider observations of a diffuse radiation field by observers comoving
with a timelike four-velocity field $u^a$. We adopt such a Lagrangian
(or 1+3 covariant) approach since it provides arguably the most transparent
description of the physics. For a recent review of the 1+3 covariant approach
in the context of relativistic cosmology see Ellis~\shortcite{ellis98}.
Relative to $u^a$, a photon with four-momentum $p^a$ is observed to have
frequency $\nu$ and propagation direction $e^a$ such that $p^a=\nu(u^a+e^a)$.
The photon propagation direction $e^a$ lies in the instantaneous rest space
of the observer, so that $u^a e_a=0$. We refer to geometric objects which are
perpendicular to $u^a$ on every index, such as $e^a$, as projected tensors.
Since Thomson scattering does not generate circular
polarization~\cite{chand-rad}, we need only consider linearly polarized
radiation here. The linear polarization state of radiation with frequency
$\nu$, which is propagating along $e^a$, is most conveniently described by
the symmetric linear polarization tensor $\clp_{ab}(\nu,e^c)$. The linear
polarization is a trace-free, projected tensor which is also perpendicular to
$e^a$. Following Thorne~\shortcite{thorne80}, we refer to projected tensors
which are also orthogonal to $e^a$ as being transverse, and we denote
the transverse, trace-free part of a rank-two tensor by the superscript
{\scriptsize{TT}}, so that $\clp_{ab}=[\clp_{ab}]^{\rmn{TT}}$. It is useful
to introduce the (screen) projection tensor
$\clh_{ab} \equiv g_{ab} - u_a u_b + e_a e_b$, where $g_{ab}$ is the
metric of spacetime. $\clh_{ab}$ is a transverse tensor which projects
perpendicular to $u^a$ and $e^a$. Making use of $\clh_{ab}$, the
properties of $\clp_{ab}(\nu,e^c)$ are summarised by
\begin{equation}
\clp_{ab} = \clp_{(ab)}, \quad \clh_a^c \clp_{bc} = \clp_{ab}, \quad
\clh^{ab} \clp_{ab} = 0,
\end{equation}
where round brackets denote symmetrisation on the enclosed indices.
The linear polarization tensor is related to
the Stokes parameters $Q(\nu,e^c)$ and $U(\nu,e^c)$ via
\begin{equation}
\clp_{11} = Q/2, \quad \clp_{12} = \clp_{21} = U/2, \quad \clp_{22} = -Q/2,
\end{equation}
where the components of $\clp_{ab}$ are on a comoving orthonormal tetrad with
the $3$-direction aligned with $e^a$.

The generation of polarization through electron scattering is most simply
described in the original rest-frame of the electron. Since we shall
ultimately give our results in the frame in which the CMB dipole vanishes,
we will need the transformation laws for the radiation variables under changes
of the velocity field $u^a$. If we consider an alternative four-velocity
field $\tilde{u}^a$, this can be written in terms of observable quantities
in the $u^a$ frame as $\tilde{u}^a=\gamma(u^a+v^a)$. Here $v^a$ is a projected
(with respect to $u^a$) four-vector which describes the relative velocity of
an observer comoving with $\tilde{u}^a$ with respect to the $u^a$ frame.
The quantity $\gamma$ is the Lorentz factor associated with the relative
motion, $\gamma=\tilde{u}^a u_a$ or equivalently $\gamma^{-2}=1+v^a v_a$.
Similarly, we can express $u^a$ in terms of quantities in the $\tilde{u}^a$
frame, so that $u^a=\gamma(\tilde{u}^a + \tilde{v}^a)$, where $\tilde{v}^a$
is the relative velocity in the $\tilde{u}^a$ frame. Introducing the
projection tensor $h_{ab}=g_{ab}-u_au_b$ in the $u^a$ frame, and a similar
projection in the $\tilde{u}^a$ frame, $\tilde{h}_{ab}$, we have
$\tilde{v}^a=-\tilde{h}^a_b v^b/\gamma$. If we consider a given photon
with four-momentum $p^a$, the frequency $\tilde{\nu}$ and propagation
direction $\tilde{e}^a$ observed in the $\tilde{u}^a$ frame are expressed
in terms of the quantities in the $u^a$ frame via the Doppler and
aberration formulae:
\begin{eqnarray}
\tilde{\nu} & = & \gamma \nu (1+ e^a v_a), \\
\tilde{e}^a & = & [\gamma(1+e^b v_b)]^{-1} (u^a + e^a) - \gamma (u^a + v^a).
\end{eqnarray}
For a given photon four-momentum $p^a$, the screen projection tensor in the
$\tilde{u}^a$ frame, $\tilde{\clh}_{ab}$, is related to $\clh_{ab}$ by
\begin{eqnarray}
\tilde{\clh}_{ab} & = & \clh_{ab} - 2 [\nu(1+e^c v_c)]^{-1} p_{(a}\clh_{b)d}
v^d \nonumber \\
&& \mbox{} + [\nu(1+e^c v_c)]^{-2} p_a p_b \clh_{d_1 d_2} v^{d_1} v^{d_2},
\end{eqnarray}
while the linear polarization tensor in the $\tilde{u}^a$ frame,
$\tilde{\clp}_{ab}(\tilde{\nu},\tilde{e}^c)$, can be expressed in terms of
the linear polarization in the $u^a$ frame as
\begin{equation}
\tilde{\nu}^{-3} \tilde{\clp}_{ab}(\tilde{\nu},\tilde{e}^c) =
\nu^{-3} \tilde{\clh}_a^{d_1} \tilde{\clh}_b^{d_2} \clp_{d_1 d_2}(\nu,e^c).
\end{equation}
This transformation law ensures that the degree of polarization
$\clp(\nu,e^c) = \surd(2\clp_{ab}\clp^{ab})/I(\nu,e^c)$, where $I(\nu,e^c)$
is the radiation intensity, is independent of frame for a given photon
momentum.

The dynamics of the radiation is described by the collisional Boltzmann
equation, which for the linear polarization takes the form
\begin{equation}
\cll [\nu^{-3}\clp_{ab}(\nu,e^c)] = K_{ab}(\nu,e^c),
\end{equation}
where the Liouville operator acts along the photon path $x^a(\lambda)$,
$p^a(\lambda)$ in phase space, with $p^a = \ud x^a / \ud \lambda$,
according to
\begin{equation}
\cll[\clp_{ab}(\nu,e^c)] = \clh_a^{d_1} \clh_b^{d_2} p^e \nabla_e 
\clp_{d_1 d_2}(\nu,e^c),
\end{equation}
where $\nabla_a$ is the spacetime covariant derivative, and
$K_{ab}(\nu,e^c)$ describes the scattering. For CMB photons scattering off
free electrons in hot clusters, the average photon energy in the rest frame
of the scattering electron is small compared to the electron rest mass,
so that electron recoil can be neglected. Furthermore, the effects of induced
scattering and Pauli blocking are negligible, so that the electron-radiation
interaction can be approximated by classical Thomson scattering. In this limit,
the scattering term in the rest frame of a beam of electrons (which has
four-velocity $\tilde{u}^a$) with number
density $\nelect$, is given by (Challinor 1999)
\begin{eqnarray}
\tilde{\nu}^2 \tilde{K}_{ab}(\tilde{\nu},\tilde{e}^c) & = &
\nelect \sigt \Bigl\{-\tilde{\clp}_{ab}(\tilde{\nu},\tilde{e}^c) \nonumber \\
&&\mbox{} + \frac{1}{10} [\tilde{I}_{ab}(\tilde{\nu})]^{\rmn{TT}}
+ \frac{3}{5} [\tilde{\cle}_{ab} (\tilde{\nu})]^{\rmn{TT}} \Bigr\},
\label{eq_k}
\end{eqnarray}
where $\sigt$ is the Thomson cross section.
Here, $I_{ab}(\nu)$ and $\cle_{ab}(\nu)$ are projected (with respect to some
general $u^a$) symmetric, trace-free (PSTF) tensors which describe the
intensity and electric polarization quadrupoles respectively in a covariant
manner~\cite{chall99}. In equation~(\ref{eq_k}), the tildes on
$\tilde{I}_{ab}$ and $\tilde{\cle}_{ab}$ denote that the intensity and
polarization quadrupoles are evaluated in the rest-frame of the electron beam,
and the transverse trace-free operations should be understood to refer
to the $\tilde{u}^a$ frame also.
For general $\ell$, we define PSTF intensity multipoles
\begin{equation}
I_{A_\ell}(\nu) = \frac{1}{\Delta_\ell} \int \ud \Omega e_{\langle
A_\ell \rangle} I(\nu,e^c),
\label{eq_int}
\end{equation}
where $\Delta_\ell \equiv 4\pi (-2)^\ell (\ell!)^2 / (2\ell+1)!$ and
$\ud\Omega$ is the element of solid angle of $e^a$, so that
\begin{equation}
I(\nu,e^c) = \sum_{\ell=0}^{\infty} I_{A_\ell}(\nu) e^{A_\ell},
\end{equation}
where $A_\ell$ is shorthand for the index string $a_1 \dots a_\ell$
and $e_{A_\ell} \equiv e_{a_1} \dots e_{a_\ell}$. Angle brackets denote the
PSTF part on the enclosed indices.
The electric polarization multipoles are defined by
\begin{equation}
\cle_{A_\ell}(\nu) = \frac{1}{\Delta_\ell}
\frac{2\ell(\ell-1)}{(\ell+1)(\ell+2)}
\int \ud \Omega e_{\langle A_{\ell-2}} \clp_{a_{\ell-1} a_\ell \rangle}
(\nu,e^c),
\end{equation}
for $\ell \geq 2$. A complete multipolar description of the linear
polarization requires the introduction of magnetic multipoles
also~\cite{kamion97,seljak97,chall99}, but we shall not need their definition
here.

In this paper we shall only be concerned with the polarization induced by
the cluster in the limit of low optical depth. Furthermore, we shall ignore
the effects of any primordial polarization since these will be small compared
to the effects of the intensity quadrupole. With these restrictions,
only the intensity quadrupole prior to the interaction with the cluster
need be retained on the right-hand side of equation~(\ref{eq_k}).
Transforming to the frame of an observer whose
four-velocity is $u^a$, the scattering term becomes
\begin{equation}
K_{ab}(\nu,e^c) = \frac{1}{10} \nelect \sigt \tilde{\nu}^{-2}
\clh_a^{d_1} \clh_b^{d_2} [\tilde{I}_{d_1 d_2}(\tilde{\nu})]^{\rmn{TT}},
\label{eq_ksim}
\end{equation}
where the transverse trace-free operation refers to the $\tilde{u}^a$
frame, and we have used the transformation law for the scattering term:
\begin{equation}
K_{ab}(\nu,e^c) = \clh_a^{d_1} \clh_b^{d_2} \tilde{K}_{d_1 d_2}(\tilde{\nu},
\tilde{e}^c).
\end{equation}
We shall compute the scattering term in the CMB frame, which has four-velocity
$u^a$, so that the intensity dipole $I_a(\nu)$ vanishes.
First we calculate the intensity quadrupole
in the rest frame of a beam of electrons, in terms of the multipoles
$I_{A_\ell}(\nu)$ in the CMB frame. Making use of the invariance of
$I(\nu,e^c)/\nu^3$, we have that
\begin{eqnarray}
\tilde{I}_{ab}(\tilde{\nu}) & = & \frac{15}{8\pi}\sum_{\ell=0}^\infty
\int \ud \tilde{\Omega} \Bigl\{ \frac{I_{C_\ell}[\gamma \tilde{\nu}
(1+\tilde{e}^d \tilde{v}_d)]}{[\gamma(1+\tilde{e}^d\tilde{v}_d)]^{(\ell+3)}}
\nonumber \\
&& \mbox{} \times (\tilde{e}^{c_1} - \tilde{v}^{c_1}) \dots (\tilde{e}^{c_\ell}
- \tilde{v}^{c_\ell}) \tilde{e}_{\langle a} \tilde{e}_{b \rangle} \Bigr\},
\label{eq_quad}
\end{eqnarray}
where $\tilde{v}^a$ is the relative velocity of $u^a$ with respect to
$\tilde{u}^a$, so that $u^a = \gamma(\tilde{u}^a + \tilde{v}^a)$, and
the integral is over solid angles of the projected directions $\tilde{e}^a$
in the $\tilde{u}^a$ frame.
Next, we use this result for $\tilde{I}_{ab}(\tilde{\nu})$ in
equation~(\ref{eq_ksim}) to obtain the scattering term in the CMB frame
due to a beam of electrons with number density $\ud \nelect$ in its rest
frame, and with velocity $v^a=-h^a_b\tilde{v}^b/\gamma$ relative to the CMB.
Finally, we express
$\ud \nelect$ in terms of the number density of the beam in the CMB frame,
using $\ud\nelec=\gamma\ud\nelect$, and integrate over the electron
distribution. Assuming a non-degenerate,
thermal distribution of electrons at temperature $\te$ in the frame
in which the electron bulk velocity vanishes, we have, in the CMB frame,
\begin{equation}
\ud \nelec = \frac{\nelec \ud^3 p_{\rmn e}}{4\pi \me^3 \thte K_2(1/\thte)}
\exp[-\gamma \Gamma (1+V^a v_a)/\thte],
\end{equation}
where $V^a$ is the projected, relative bulk velocity, $\Gamma$
is the associated Lorentz factor, and $\ud^3 p_{\rmn e}$
is the element of 3-momentum. Here, $\thte \equiv \kb\te/\me$ is the
dimensionless electron temperature, $\me$ is the electron mass,
$K_2(x)$ is a modified Bessel function, and $\nelec$ is the electron number
density in the frame in which the bulk velocity vanishes. Note that the
electron number density in the CMB frame is $\Gamma \nelec$.

\section{Polarization from the multipoles $I_{A_\ell}(\nu)$}
\label{results}

In this section we compute the contribution to the scattering term
$K_{ab}(\nu, e^c)$ from each primordial multipole $I_{A_\ell}(\nu)$ in turn,
expressing our results as expansions in the small parameters
$\thte= \kb\te/(511\,\rmn{keV})$ and $V \equiv \sqrt{-V^a V_a}$.

In linear perturbation theory, where there is no non-linear scattering,
there are no spectral distortions in the primordial radiation, so that
\begin{equation}
I(\nu,e^c) = \frac{2\nu^3}{{\rmn{e}}^{\nu/[\kb\bar{T}_{\rmn{r}}(1+\delta_T)]}
-1},
\end{equation}
where $\bar{T}_{\rmn{r}}$ is the all-sky average CMB temperature, and
$\delta_T(e^c)$ is the dimensionless temperature anisotropy. If we expand
$\delta_T(e^c)$ in PSTF multipoles $\tau_{A_\ell}$~\cite{maartens95},
\begin{equation}
\delta_T(e^c) = \sum_{\ell=2}^{\infty} \tau_{A_\ell} e^{A_\ell},
\end{equation}
(recall there is no dipole in the CMB frame), then, to first-order in
the primordial anisotropy, we have for $\ell>1$
\begin{equation}
I_{A_\ell}(\nu) = i(x) \tau_{A_\ell}, \quad {\rmn{where}}, \quad
i(x) \equiv I(\nu) \frac{x{\rmn{e}}^x}{{\rmn{e}}^x-1},
\end{equation}
with $I(\nu)$ the intensity monopole (a Planck spectrum with
temperature $\bar{T}_{\rmn{r}}$), and $x\equiv \nu/(\kb\bar{T}_{\rmn{r}})$
the dimensionless frequency. In this manner, the frequency dependence of the
intensity multipoles is contained in the scalar $i(x)$ only, which proves to
be very convenient for subsequent calculations. It should be noted that since
$\delta_T(e^c)$ has no monopole component by definition, no confusion should
arise from our later use of the symbol $\tau$ for the optical depth through
the cluster.

\subsection{The monopole contribution}

The contribution to the intensity quadrupole in the rest frame of the
electron beam from the intensity monopole in the CMB frame, $I(\nu)$,
follows from equation~(\ref{eq_quad}):
\begin{equation}
\tilde{I}_{ab}(\tilde{\nu}) = \frac{15}{8\pi}\int\ud\tilde{\Omega}
\frac{I[\gamma\tilde{\nu}(1+\tilde{e}^c\tilde{v}_c)]}{[\gamma(1+\tilde{e}^c
\tilde{v}_c)]^3}\tilde{e}_{\langle a}\tilde{e}_{b\rangle},
\end{equation}
which evaluates to
\begin{equation}
\tilde{I}_{ab}(\tilde{\nu}) = \frac{15}{8\beta^2}\int_{-1}^{+1}
\ud x\frac{I[\gamma\tilde{\nu}(1-\beta x)]}
{[\gamma(1-\beta x)]^3}(3x^2-1) \tilde{v}_{\langle a}\tilde{v}_{b\rangle},
\label{eq_mon}
\end{equation}
where $\beta^2 = - \tilde{v}^a \tilde{v}_a = - v^a v_a$. Note that the
contribution to $\tilde{I}_{ab}(\tilde{\nu})$ from the CMB monopole
is $O(\beta^2)$. To transform to the CMB frame, we require the result
\begin{equation}
\clh_a^{c_1} \clh_b^{c_2} [\tilde{v}_{\langle c_1} \tilde{v}_{c_2 \rangle}
]^{\rmn{TT}} = [\gamma(1+e^c v_c)]^{-2} [v_{\langle a}v_{b\rangle}]^{\rmn{TT}},
\label{eq_montrans}
\end{equation}
which follows from elementary manipulations. The transverse trace-free and PSTF
operations on the left-hand side of equation~(\ref{eq_montrans}) refer to
the $\tilde{u}^a$ frame, while those on the right refer to the $u^a$ frame.
It follows that
the scattering term $K_{ab}(\nu,e^c)$ due to a beam of electrons with
relative velocity $v^a$ is proportional to $[v_{\langle a}
v_{b\rangle}]^{\rmn{TT}}$, so that the radiation is polarized along the
normal to the plane containing $v^a$ and the line of sight $e^a$.

We use equations~(\ref{eq_mon}) and~(\ref{eq_montrans}) in the right-hand side
of equation~(\ref{eq_ksim}), and expand in powers of $\beta$. The
integral over the electron distribution is handled by an expansion
in powers of $V$ and $\thte$. The integrals are straightforward, but tedious,
making them ideally suited to symbolic computer algebra packages
(we use \textsc{maple}).
The final result is an asymptotic expansion for the scattering
term in the CMB frame due to the intensity monopole, of which the leading
few terms are
\begin{eqnarray}
\nu^2 K_{ab}(\nu,e^c) & = & \frac{1}{10}\nelec\sigt [V_{\langle a}
V_{b\rangle}]^{\rmn{TT}} I(\nu) \frac{x {\rmn{e}}^x}{{\rmn{e}}^x-1}
\Bigl\{\frac{1}{2}F \nonumber \\
&& \mbox{} + \thte\Bigl[3F - 2(2F^2+G^2) \nonumber \\
&& \mbox{} +\frac{1}{2}F(F^2+2G^2)\Bigr] \nonumber \\
&& \mbox{} + V \mu \left[-\frac{1}{2}F + \frac{1}{4}(2F^2+G^2)\right] \Bigr\},
\label{eq_monres}
\end{eqnarray}
where $\mu \equiv -V^a e_a /V$ is the cosine of the angle between the
bulk velocity of the cluster and the photon propagation direction.
Here, $F\equiv x \coth(x/2)$ and $G\equiv x/\sinh(x/2)$. The presence of the
factor $[V_{\langle a}V_{b\rangle}]^{\rmn{TT}}$ ensures that the
radiation is polarized along the normal to the plane containing the
bulk velocity and the line of sight, as required by symmetry considerations.

If we assume that the distance across the cluster is small compared to the
characteristic length scale of the local spacetime curvature, then in the
limit of low optical depth the emergent polarization is simply
$\clp_{ab}(\nu,e^c)=\nu^2 K_{ab}(\nu,e^c) t$, where $t$ is the light
travel time across the cluster in the CMB frame. To obtain the degree
of polarization $\clp(\nu,e^c)$ from equation~(\ref{eq_monres}), one
should multiply by $t$, take the magnitude (times $\sqrt{2}$), and divide
by the emergent intensity $I(\nu,e^c)+\Delta I(\nu,e^c)$, where
$\Delta I(\nu,e^c)$ is the change in intensity due to the scattering.
Ignoring the small anisotropy in the CMB frame, the
leading order terms of $\Delta I(\nu,e^c)$
are~\cite{sunyaev72,sunyaev80}
\begin{equation}
\Delta I(\nu,e^c)=\nelec \sigt t I(\nu) \frac{x {\rmn{e}}^x}{{\rmn{e}}^x-1}
\left[\thte(F-4)+V\mu\right].
\end{equation}
Higher order terms in the expansion of $\Delta I(\nu,e^c)$ have been
calculated
recently~\cite{stebbins97,chall98,chall99a,itoh98,nozawa98,sazonov98}.
Identifying the effective optical depth through the cluster,
$\tau\equiv\nelec\sigt t$, in the limit of small $\thte$ and $V$ we recover
the result for $\clp(\nu,e^c)$ given as equation (22) in
Sazonov \&\ Sunyaev~\shortcite{sazonov99}:
\begin{equation}
\clp(\nu,e^c)=\frac{x^2 {\rmn{e}}^x}{20({\rmn{e}}^x-1)}
\coth(x/2) \tau \vt^2,
\label{eq_saz}
\end{equation}
where we have used
\begin{equation}
2[V_{\langle a}V_{b\rangle}]^{\rmn TT} [V^{\langle a}V^{b\rangle}]^{\rmn TT}
= \vt^4,
\end{equation}
with $\vt^2 = V^2 (1-\mu^2)$ the square of the transverse velocity.

The $O(\thte^0,V^0)$ and $O(\thte^1,V^0)$ terms in the curly braces
on the right-hand side of equation~(\ref{eq_monres}) correct those
given by Itoh et al.~\shortcite{itoh99}. Itoh et al.\ also give an expression
for the $O(\thte^2,V^0)$ term; we include our result here for completeness:
\begin{eqnarray}
&&\mbox{}\thte^2\Bigl\{\frac{15}{2}F - 21(2F^2+G^2)+\frac{867}{28} F(F^2+2G^2)
\nonumber \\
&& \mbox{} -\frac{40}{7}\left[F^4 + \frac{1}{2}G^2(11F^2+2G^2)\right]
\nonumber \\
&& \mbox{} +\frac{2}{7}F\left[F^4+\frac{1}{2}G^2(26F^2+17G^2)\right] \Bigr\}.
\label{eq_monres2}
\end{eqnarray}
In the Rayleigh-Jeans region, neglecting $\Delta I(\nu,e^c)$ compared to
$I(\nu,e^c)\approx I(\nu)$, we find the degree of polarization
\begin{equation}
\clp(\nu,e^c) = \frac{1}{10}\tau \vt^2\left(1-6\thte+2V\mu + \frac{183}{7}
\thte^2 + \dots\right),
\label{eq_monrj}
\end{equation}
which, as noted by Sazonov \&\ Sunyaev~\shortcite{sazonov99},
recovers the result of
Sunyaev \&\ Zel'dovich~\shortcite{sunyaev80} at lowest order. The
$O(\thte^0,V^1)$ term in the braces on the right of equation~(\ref{eq_monres})
is responsible for the weak dependence of the degree of polarization
on the absolute value of the bulk velocity, for a given transverse
velocity $\vt$, as found by Audit \&\ Simmons~\shortcite{audit99} through
numerical calculations in the zero temperature limit.

\begin{figure}
\epsfig{figure=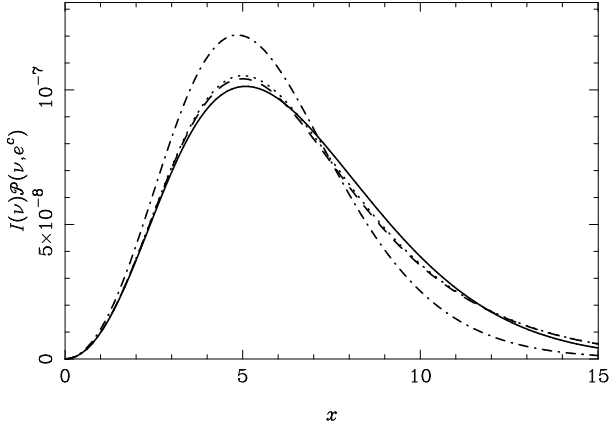,angle=-90,width=8cm}
\caption{Magnitude of the polarization [in units of $2(\kb\bar{T}_{\rmn{r}}
)^3/(hc)^2$] due to the CMB monopole for a cluster with  $\kb\te =
10\,{\rmn{keV}}$,
$\vt=1000\,{\rmn{kms^{-1}}}$, $\tau = 0.01$, and $\mu=1/\sqrt{3}$.
The dashed-dotted line is the contribution with only the $O(\thte^0,V^0)$ term
in curly braces in equation~(\ref{eq_monres}) included.
The solid line also includes the $O(\thte^1,V^0)$ term, and the dashed line
further includes the $O(\thte^2,V^0)$ term given as
equation~(\ref{eq_monres2}). The dotted line includes these first two thermal
corrections, as well as the $O(\thte^0,V^1)$ term.}
\label{fig1}
\end{figure}

In Fig.~\ref{fig1} we plot the frequency dependence of the
magnitude of the polarization, $I(\nu) \clp(\nu,e^c)$, in units
of $2(\kb\tr)^3/(hc)^2$, for the
cluster parameters $\kb\te = 10\,{\rmn{keV}}$, $\vt=1000\,{\rmn{kms^{-1}}}$,
$\tau = 0.01$. The direction of the bulk motion is chosen to maximise the
$O(\thte^0,V^1)$ term in equation~(\ref{eq_monres}), so that
$\mu=1/\sqrt{3}$. For these parameters, the leading order
correction to the polarization from the CMB monopole is thermal in origin,
and tends to broaden the polarization signal.
In the Rayleigh-Jeans region the correction amounts to a reduction in
the signal of $\simeq 10$ per cent,
which also follows from equation~(\ref{eq_monrj}),
while near the peak of the polarized signal ($x=4.85$) the reduction is
$\simeq 15$ per cent. Near the peak, including the $O(\thte^2,V^0)$
term in curly braces in equation~(\ref{eq_monres}) gives a further
$3$ per cent correction upwards, while including the
$O(\thte^0,V^0)$ term (the leading order kinematic correction) gives
a further correction upwards of only $1$ per cent when the radial motion is
towards the cluster. In the Wien region, where the degree of polarization
is large~\cite{audit99}, thermal effects lead to a much larger fractional
enhancement in the polarization. For example, in the \emph{Planck} HFI
channel at $545\,{\rmn{GHz}}$, the thermal enhancement of the
polarization due to the monopole is nearly 30 per cent. However,
for all of the \emph{Planck} channels, the degree of polarization from the
cluster parameters in Fig.~\ref{fig1} falls well below the predicted detector
sensitivity. In the \emph{Planck} HFI channels at $143$,
$217$ and $545\,{\rmn{GHz}}$, the magnitude of the polarization
from the monopole as an equivalent fractional temperature change
$\delta T/T$ is $\simeq 0.014$, $0.019$, and $0.070\,\umu {\rmn{K}}/{\rmn{K}}$
respectively. These values should be compared to
the predicted sensitivities of $3.7$, $8.9$ and $208\,
\umu{\rmn{K}}/{\rmn{K}}$. Our conclusions concerning the feasibility of
direct detection of cluster polarization with the \emph{Planck} satellite
are less optimistic than those in Audit \& Simmons~\shortcite{audit99}.
This is due to our use of rather more realistic cluster parameters,
and the correct predicted equivalent thermodynamic temperature sensitivities.

The errors in our analytic expressions for the polarization (and also
the intensity change) decrease with decreasing cluster temperature and
bulk velocity. For fixed parameters, the errors increase with increasing
frequency. The latter behaviour is due to the use of a Fokker-Planck expansion
of the intensity with respect to the frequency change on scattering in the CMB
frame, during the evaluation of $\tilde{I}_{ab}(\tilde{\nu})$
[equation~(\ref{eq_quad})]. In practice this limits the range of frequencies
for which the expansions are valid to $x \la 1/\beta$ (recall that
$\beta$ is the speed of the scattering electron relative to the CMB).
For a typical cluster, the thermal electron motion is the limiting factor,
giving a maximum dimensionless frequency $x$ of the order of 10.
Note that equation~(\ref{eq_saz}) suggests a degree of polarization which
behaves as $x^2$ for large $x$. However, this is not in conflict with the 
requirement that $\clp(\nu,e^c)$ be bounded by unity given the limited range
in $x$ for which equation~(\ref{eq_saz}) is valid.

\subsection{The quadrupole contribution}

The contribution to the quadrupole in the rest frame of a beam of electrons
from the primordial quadrupole in the CMB frame is given by
\begin{eqnarray}
\tilde{I}_{ab}(\tilde{\nu}) & = & \frac{15}{8\pi}
\int \ud \tilde{\Omega} \Bigl\{ \frac{i[\gamma \tilde{\nu}
(1+\tilde{e}^d \tilde{v}_d)]}{[\gamma(1+\tilde{e}^d\tilde{v}_d)]^5}
\nonumber \\
&& \mbox{} \times \tau_{c_1 c_2}(\tilde{e}^{c_1} - \tilde{v}^{c_1})
(\tilde{e}^{c_2} - \tilde{v}^{c_2}) \tilde{e}_{\langle a}
\tilde{e}_{b \rangle} \Bigr\}.
\end{eqnarray}
Evaluating this integral is a little more involved than for the monopole
calculation given in the previous subsection. An expansion in powers of
$\beta$ gives three distinct sets of terms. First, there are terms
proportional to $\tilde{h}^{c_1}_{\langle a}\tilde{h}^{c_2}_{b\rangle}
\tau_{c_1 c_2}$, where it will be recalled that $\tilde{h}_{ab}
\equiv g_{ab} - \tilde{u}_a \tilde{u}_b$ is the projection tensor into the
instantaneous rest space of the $\tilde{u}^a$ frame.
These terms transform to the CMB frame according to
\begin{eqnarray}
\clh^{c_1}_a \clh^{c_2}_b [\tilde{h}^{d_1}_{\langle c_1}
\tilde{h}^{d_2}_{c_2 \rangle}\tau_{d_1 d_2}]^{\rmn{TT}} & = &
[\tau_{ab}]^{\rmn{TT}} - \frac{2}{\alpha}[v_{\langle a}\tau_{b\rangle c}
e^c]^{\rmn{TT}} \nonumber \\
&&\mbox{} \hspace{-0.5cm}+ \frac{1}{\alpha^2} \tau_{c_1 c_2} e^{c_1}e^{c_2}
[v_{\langle a} v_{b\rangle}]^{\rmn{TT}},
\label{eq_quadtrans}
\end{eqnarray}
where $\alpha\equiv 1+e^c v_c$. Second, there are terms proportional to
$\tilde{v}_{\langle a} \tilde{h}^{c_1}_{b\rangle} \tilde{v}^{c_2}
\tau_{c_1 c_2}$, which transform as
\begin{eqnarray}
\clh^{c_1}_a \clh^{c_2}_b [\tilde{v}_{\langle c_1} \tilde{h}^{d_1}_{c_2\rangle}
\tilde{v}^{d_2} \tau_{d_1 d_2}]^{\rmn{TT}} & = & 
- \frac{1}{\gamma\alpha}[v_{\langle a}\tau_{b\rangle c}v^c]^{\rmn{TT}}
\nonumber \\
&&\mbox{} \hspace{-1.5cm}+ \frac{1}{\gamma \alpha^2}
[v_{\langle a}v_{b\rangle}]^{\rmn{TT}} \tau_{c_1 c_2} v^{c_1} e^{c_2}.
\label{eq_quadtransb}
\end{eqnarray}
In equations~(\ref{eq_quadtrans}) and (\ref{eq_quadtransb}) the
transverse trace-free and PSTF operations on the left-hand sides refer to the
$\tilde{u}^a$ frame, while those on the right refer to the $u^a$ frame.
Finally, there are terms proportional to $\tilde{v}_{\langle a}
\tilde{v}_{b\rangle} v^{c_1} v^{c_2} \tau_{c_1 c_2}$, whose transformation law
follows from equation~(\ref{eq_montrans}). It follows that, in general, the
polarization due to the primordial quadrupole will be a superposition of
six geometric objects: the five terms on the right-hand sides of
equations~(\ref{eq_quadtrans}) and~(\ref{eq_quadtransb}), with
$v^a$ replaced by the cluster peculiar velocity $V^a$, and a term
$V_{\langle a}V_{b\rangle} V^{c_1} V^{c_2} \tau_{c_1 c_2}$.
If the peculiar velocity of the cluster vanishes, the polarization tensor
is proportional to $[\tau_{ab}]^{\rmn{TT}}$ only. The latter case was discussed
by Sazonov \&\ Sunyaev~\shortcite{sazonov99} in the low temperature limit;
it is straightforward to show that the dependence of the degree of polarization
on the position of the cluster on the sky, for $\clp_{ab}(\nu,e^c)\propto
[\tau_{ab}]^{\rmn{TT}}$, is equivalent to that derived in
Sazonov \&\ Sunyaev~\shortcite{sazonov99} by non-geometric means.

The leading few terms for the scattering term in the CMB frame due to the
primordial quadrupole are
\begin{eqnarray}
\nu^2 K_{ab}(\nu,e^c) & = & 
\frac{1}{10}\nelec\sigt I(\nu) \frac{x {\rmn{e}}^x}{{\rmn{e}}^x-1}
\Bigl( [\tau_{ab}]^{\rmn{TT}}\Bigl\{1+\thte \bigl[
-6F \nonumber \\
&&\mbox{} +\frac{1}{2}(2F^2+G^2)\bigr] + V\mu(-2+F) \Bigr\}
\nonumber \\
&&\mbox{} - 2[V_{\langle a}\tau_{b\rangle c}e^c]^{\rmn{TT}} \Bigr).
\label{eq_quadres}
\end{eqnarray}
For completeness, we also give the $O(\thte^2)$ term. This only has a
$[\tau_{ab}]^{TT}$ geometric dependence, and so would appear in the curly
braces in equation~(\ref{eq_quadres}):
\begin{eqnarray}
&& \mbox{}\thte^2\Bigl\{-15F+\frac{69}{4}(2F^2+G^2)
- \frac{64}{7}F(F^2+2G^2)\nonumber \\
&& \mbox{} +\frac{4}{7}\left[F^4+\frac{1}{2}G^2(11F^2+2G^2)
\right] \Bigr\}.
\label{eq_quadres2}
\end{eqnarray}

In the limit of small $\thte$ and $V$, the polarization is given by
\begin{equation}
\clp_{ab}(\nu,e^c) = \frac{1}{10}\tau i(x) [\tau_{ab}]^{\rmn{TT}}
= \frac{1}{10} \tau [I_{ab}(\nu)]^{\rmn{TT}},
\end{equation}
which also follows directly from equation~(\ref{eq_ksim}) on setting
$v^a=0$. This very simple result is equivalent to that derived by
Sazonov \&\ Sunyaev~\shortcite{sazonov99}, but the presentation here benefits
from the fully geometric approach. We can write $\tau_{ab}$ in the form
$\bar{T}_r \tau_{ab} = \sum_{i=1}^3 \lambda_i n^{(i)}_a n^{(i)}_b$, where
$\{\lambda_i\}$ are the (observed) eigenvalues of the temperature
quadrupole, and $\{n^{(i)}_a\}$ are the corresponding unit projected
eigenvectors. (For the metric signature employed here,
$\bar{T}_r {\tau_a}^b n^{(i)}_b = - \lambda_i n^{(i)}_a$.) For
$\clp_{ab}(\nu,e^c)\propto [\tau_{ab}]^{\rmn{TT}}$, it follows that the
degree of polarization is maximal along the eigendirection with
intermediate eigenvalue (say $\lambda_3$). Perpendicular to this direction,
the polarization vanishes for directions such that
$(e^a n^{(1)}_a)^2 = (\lambda_1-\lambda_3)/(\lambda_1-\lambda_2)$.
For clusters at these positions on the sky, the polarization signal due to the
primordial quadrupole will vanish except for small corrections arising from
the cluster peculiar velocity. The first such correction is given in
equation~(\ref{eq_quadres}). Including only this correction, the
degree of polarization due to the primordial quadrupole
along directions for which $[\tau_{ab}]^{\rmn{TT}}$ vanishes evaluates to
\begin{equation}
\clp = \frac{1}{10}\tau \vt 2 [(\lambda_3-\lambda_1)(\lambda_2-\lambda_3)
]^{1/2},
\end{equation}
which we have expressed as an equivalent temperature.
In the Rayleigh-Jeans region, we find
\begin{eqnarray}
\clp_{ab}(\nu,e^c) &=& \frac{1}{10}\tau I(\nu)
\Bigl[[\tau_{ab}]^{\rmn{TT}}\left(1-6\thte
+\frac{183}{7}\thte^2 \right)  \nonumber \\
&&\mbox{}- 2 [V_{\langle a}\tau_{b \rangle c}e^c]^{\rmn{TT}} \Bigr].
\label{eq_quadrj}
\end{eqnarray}
Note that the $O(V)$ correction to the $[\tau_{ab}]^{\rmn{TT}}$
polarization term vanishes in this limit.

\begin{figure}
\epsfig{figure=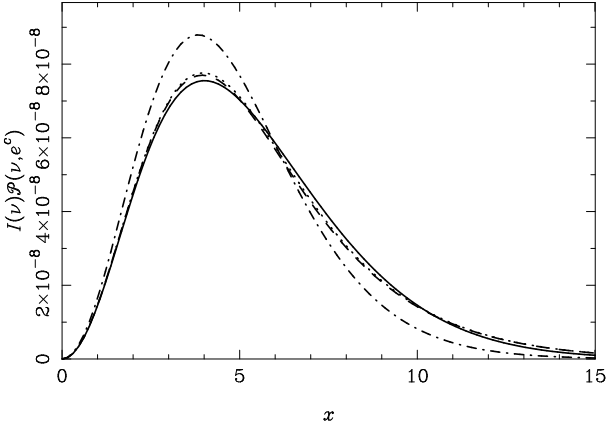,angle=-90,width=8cm}
\caption{Magnitude of the dominant $[\tau_{ab}]^{\rmn{TT}}$
polarization [in units of $2(\kb\bar{T}_{\rmn{r}})^3/(hc)^2$]
due to the CMB quadrupole for a cluster with  $\kb\te = 10\,{\rmn{keV}}$,
$V\mu=1000\,{\rmn{kms^{-1}}}$ and $\tau = 0.01$. We use the best-fitting
\emph{COBE} results for the quadrupole: $\lambda_1=-25\,\umu\rmn{K}$,
$\lambda_2=23\,\umu\rmn{K}$ and $\lambda_3=2\,\umu\rmn{K}$.
The dashed-dotted line is the contribution with only the $O(\thte^0,V^0)$ term
in curly braces in equation~(\ref{eq_quadres}) included.
The solid line also includes the $O(\thte^1,V^0)$ term, and the dashed line
further includes the $O(\thte^2,V^0)$ term given as
equation~(\ref{eq_quadres2}). The dotted line includes these first two thermal
corrections, as well as the $O(\thte^0,V^1)$ term.}
\label{fig2}
\end{figure}

In Fig.~\ref{fig2} we plot the magnitude of the dominant polarization
signal (that with $[\tau_{ab}]^{\rmn{TT}}$ geometric dependence) due
to the CMB quadrupole, for the cluster parameters $\kb\te=10\,\rmn{keV}$,
$V\mu=1000\,\rmn{kms^{-1}}$ and $\tau=0.01$. For the CMB quadrupole we
take the \emph{COBE} best-fitting values  $\lambda_1=-25\,\umu\rmn{K}$,
$\lambda_2=23\,\umu\rmn{K}$ and $\lambda_3=2\,\umu\rmn{K}$ used by
Sazonov \&\ Sunyaev~\shortcite{sazonov99}. The line of sight is chosen to
lie along $n^{(3)}_a$ to maximise the polarization signal.
As with the polarization due to the
CMB monopole, we see that the dominant correction to the polarization
sourced by the primordial quadrupole is due to thermal effects, and these tend
to broaden the signal. In the Rayleigh Jeans region the size of the fractional
correction is the same as for the signal due to the monopole $\simeq 10$
per cent [compare equations~(\ref{eq_monrj}) and~(\ref{eq_quadrj})].
Near the signal peak ($x=3.83$) the leading thermal correction leads to a
reduction of $\simeq 14$ per cent, but this is reduced to $\simeq 12$
per cent when the next order thermal correction and the $O(V)$ kinematic
term are included. The corrections are most significant in
the Wien region; in the $545\,\rmn{GHz}$ \emph{Planck}
HFI channel the thermal terms increase the signal by $\simeq 60$ percent.
However, since the monopole contribution to the degree of polarization
rises more rapidly with frequency (as $x^2$) than the quadrupole
(as $x$) for large $x$, the quadrupole contribution is likely to be less
significant than the monopole in the Wien region, for typical cluster
parameters~\cite{sazonov99}. In the \emph{Planck} HFI channels at $143$,
$217$ and $545\,{\rmn{GHz}}$, the magnitude of the polarization
from the quadrupole as an equivalent fractional temperature change
$\delta T/T$ is $\simeq 0.016$, $0.016$, and $0.029\,\umu {\rmn{K}}/{\rmn{K}}$
respectively, which, like the monopole contribution, is well below the
predicted sensitivities.

\subsection{The contribution of higher multipoles}

The $\ell$-th multipole of the CMB anisotropy gives rise to a quadrupole
of $O(\beta^{\ell-2})$ in the rest frame of the scattering electron.
It follows that the leading terms in the signal due to the CMB octupole
($\ell=3$) will be $O(V,\thte)$. Unlike the monopole, the octupole
can give a thermal polarization signal even if the cluster peculiar
velocity vanishes ($V=0$), due to the preferred directions in space that it
defines. Since the octupole is of similar magnitude to
the quadrupole [the \emph{COBE} best-fitting values give
$|\tau_{abc}|/|\tau_{ab}| \simeq 2.3$~\cite{stoeger97}],
and given that the $O(\thte)$ term in the quadrupole
contribution to the polarization can be a sizeable fraction of the
total quadrupole signal, the octupole contribution should potentially
be included in future attempts to determine the primordial quadrupole at the
location of distant (hot) clusters via the cluster polarization
route~\cite{kamion97b}.

The calculation of the polarization due to the CMB octupole follows that of the
monopole and quadrupole. For the leading two terms we find
\begin{eqnarray}
\nu^2 K_{ab}(\nu,e^c) & = & \frac{1}{10} \nelec
\sigt I(\nu) \frac{x {\rmn{e}}^x} {{\rmn{e}}^x-1}\frac{3}{7}\Bigl\{
[\tau_{abc}V^c]^{\rmn{TT}} (5-F) \nonumber \\
&&\mbox{}\hspace{-0.4cm}+ \thte [\tau_{abc}e^c]^{\rmn{TT}}
\left[6F - \frac{1}{2}(2F^2+G^2)\right] \Bigr\}.
\end{eqnarray}
For the hexadecapole ($\ell=4$) the quadrupole in the rest frame of the
electron is second-order in $\beta$, so we might anticipate an
$O(\thte)$ polarization signal. However, to this order the polarization from
a beam of electrons with relative velocity $v^a$ is proportional to
$[\tau_{abcd}v^c v^d]^{\rmn{TT}}$, which results in a polarization signal
proportional to $[\tau_{abcd}V^c V^d]^{\rmn{TT}}$ on integrating
over the electron distribution. For reasonable values of the cluster
peculiar velocity, we infer that the effects of the CMB multipoles
with $\ell>3$ can be safely ignored, even for rather hot clusters.

\section{Conclusion}
\label{conc}

We have presented a direct geometric method for determining the
polarization due to the scattering of anisotropic radiation off
a thermal distribution of electrons, with non-vanishing bulk
velocity, in the limit of low optical depth. Our results for the polarization
due to the CMB monopole and quadrupole fully agree with those obtained by
Sazonov \&\ Sunyaev~\shortcite{sazonov99}
[see also Zel'dovich \&\ Sunyaev~\shortcite{zeldovich80}]
in the low temperature
limit. Our results for the thermal corrections to the monopole effect correct
those given by Itoh et al.~\shortcite{itoh99}. A careful inspection of their
analysis suggests that the discrepancy arises from the omission
of a rotation needed to relate the scattering basis in the rest frame
of the initial electron and the CMB frame. We have shown that for typical
cluster parameters the thermal corrections tend to dominate the
kinematic corrections. Near the peak of their respective polarization signals,
thermal effects tend to reduce the magnitude of the polarization from the
monopole and quadrupole by $\simeq 12$ per cent. The corrections are much
larger in the Wien part of the spectrum. We have also noted that the
CMB octupole can lead to a polarization signal from hot clusters which
is of similar magnitude to the thermal correction to the quadrupole signal.
The effects reported here may be significant for future attempts to
determine the CMB quadrupole at distant clusters and the tangential
peculiar velocity, to an accuracy of a few per cent, via polarization
observations.

\section*{Acknowledgements}

ADC acknowledges Queens' College, Cambridge for support in the
form of a Research Fellowship.

\bsp  

\end{document}